\documentclass[preprint,authoryear,12pt]{elsarticle}
\usepackage{epsfig}
\usepackage{amssymb}
\usepackage[ps2pdf,%
a4paper=true,%
breaklinks=true,%
colorlinks=true,%
pdfauthor={First Author et al.},%
pdftitle={Template for manuscripts in Advances in Space Research}%
]{hyperref}
\journal{Advances in Space Research}

\begin{document}
\begin{frontmatter}
\title{World Space Observatory-Ultraviolet: \\
ISSIS, the imaging instrument \tnoteref{footnote1}}
\tnotetext[footnote1]{The World Space Observatory-Ultraviolet 
mission is described in \citet{Sac13a}.}
\author{Ana I. G\'omez de Castro\corref{cor}}
\cortext[cor]{Corresponding author}
\ead{aig@mat.ucm.es}
\author{Paola Sestito,\\
N\'estor S\'anchez, F\'atima L\'opez-Mart\'{\i}nez}
\address{AEGORA, Fac. de CC. Matem\'aticas,
Universidad Complutense de Madrid,\\
Plaza de Ciencias 3, 28040 Madrid, Spain}
\author{Juan Seijas, Maite G\'omez, Pablo Rodr\'{\i}guez,\\
Jos\'e Quintana, Marcos Ubierna, Jacinto Mu\~noz}
\address{SENER S.A., C/ Severo Ochoa 4, 28760 Tres Cantos, Madrid, Spain}
\begin{abstract}
The Imaging and Slitless Spectroscopy Instrument (ISSIS)
will be flown as part of the science instrumentation in
the World Space Observatory-Ultraviolet (WSO-UV). ISSIS
will be the first UV imager to operate in a high Earth
orbit from a 2~m class space telescope. In this contribution,
the science driving to ISSIS design, as well as main
characteristics of ISSIS are presented.
\end{abstract}
\end{frontmatter}

\parindent=0.5 cm

\section{Introduction}

The resonance transitions of the most abundant species in the
Universe as well as the electronic transitions of the most 
abundant molecules are in the ultraviolet (UV) range of the
spectrum. Hence, access to the UV range is instrumental
for progress in Astrophysics \citep{Gom06}. The World 
Space Observatory - Ultraviolet (WSO-UV) space telescope
will grant this access in the post-Hubble Space Telescope
era. WSO-UV is a 170~cm diameter primary telescope with an
efficient Ritchey-Chr\'etien configuration that will be 
operational in the 2017--2027 time frame. WSO-UV is a
Russian-led mission and will be equipped with instrumentation 
for UV spectroscopy and imaging \citep[see][]{Shu11,Sac13a}.

The Imaging and Slitless Spectroscopy Instrument (ISSIS) will
be a key part of the WSO-UV instrumentation. ISSIS is the first
UV imager to be flown to high Earth orbit, above the Earth
geocorona. Therefore, the UV background will be dominated by
the zodiacal contribution and the diffuse galactic background due
to dust-scattered starlight \citep{Suj09,Mur10}. The instrument
has been designed to make full benefit of the heritage left
by the GALactic Evolution eXplorer (GALEX) mission. GALEX has
surveyed about 80\,\% of the sky at UV wavelengths, providing
for the first time a nearly complete view of the UV Universe
\citep{Mar03,Bia11}. However, GALEX spatial resolution was
$\sim 4.2$~arcsec and had very moderate spectroscopic
capabilities. ISSIS resolution will be $\leq 0.1$~arcsec.
The Fine Guiding System of the WSO-UV telescope will
guarantee a very high pointing stability (better than
0.1~arcsec at 3$\sigma$). Moreover, ISSIS will be equipped
with gratings for slitless spectroscopy with spectral
resolution 500 in the full 1150--3200~\AA\ spectral range.
In imaging mode, the ISSIS effective area is about 10 times
that of the GALEX imagers.

ISSIS is designed to be an instrument for analysis of weak
UV point sources or clumpy extended sources, especially
those with well defined geometry. UV imaging instruments
have been often equipped with prisms or very low dispersion
grisms. The rapid decay of the resolution of prisms, such as
the available in the Solar Blind Channel of the Advanced 
Camera System, makes very difficult its use to map extended
line emission at wavelengths above some 1350~\AA . As the
transmittance of narrow band filters in the far UV is
$\leq 3$\,\%, integral-field low-resolution spectroscopy
is the main mean to map nebular emission. ISSIS gratings
will make feasible to use the powerful UV diagnostic tools
to determine the location of dusty blobs and measure electron
densities and temperatures. In this contribution, the science
that has driven to ISSIS design is described in Section~2.
ISSIS design is presented in Section~3 and its performance 
is described in Section~4. ISSIS planning and a brief summary
are provided in the concluding Section~5.

\section{Science with ISSIS}

The key scientific drivers of the WSO-UV project are:

\begin{itemize}
\item Galaxy formation: the evolution of the Star Formation
Rate with redshift ($z$) and the role of the intergalactic
medium (IGM) as a mass supply for star formation at moderate
redshifts ($z<2$) -- To study the characteristics of the
diffuse baryonic content of the Universe and its chemical
evolution; to determine, in particular, the baryonic content
of the warm and hot IGM and of damped Lyman-$\alpha$ systems;
and to investigate the role of starbursts in the chemical
evolution of galaxies and the IGM.
\item The formation and evolution of the Milky Way -- To
estimate the energy inputs of the gas interacting with
stars; and to investigate the role of magnetic fields on
star formation.
\item The physics of accretion and outflows: the astronomical
engines -- To study stars, black holes, interacting binaries
and, in general, all those objects where mass accretion has a
relevant impact in the system evolution; and to analise the
efficiency and time scales of these phenomena as well as the
role of magnetic fields, radiation pressure and disk instabilities.
\item Extrasolar planetary atmospheres and astrochemistry in
presence of strong UV radiation fields -- To measure the UV
radiation field from T~Tauri stars and its impact on the
chemical evolution of young planetary disks and planetary
atmospheres. 
\end{itemize}

For this purpose, the WSO-UV telescope is equipped with the
imaging instrument ISSIS and the three WSO-UV Spectrographs
\citep[WUVS; see][]{Sac10,Sac13b}:

\begin{itemize}
\item VUVES, the far-UV high-resolution echelle spectrograph
operating in the 1115--1760~\AA\ range with resolution R$\sim$50,000.
\item UVES, the near-UV high-resolution echelle spectrograph in the
1740--3100~\AA\ range with resolution R$\sim$50,000.
\item LSS, the Long-Slit Spectrograph (LSS) for low-resolution
(R$\sim 100$) long-slit spectroscopy. The width of the slit
will be 1~arcsec and the spatial resolution 0.5~arcsec.
\end{itemize} 

LSS and ISSIS provide together a powerful tool to study extended
sources since the high resolution imaging of ISSIS complements
the LSS field selection capability. Both ISSIS and LSS have
similar spectral resolutions.

The main topics that could be investigated with the instrument
ISSIS cover a wide range of astronomical issues:

\begin{itemize}
\item Solar System -- The atmospheres of the planets and
satellites will be investigated as well as the impact of the
Solar radiation in their evolution. The connection between
the Solar System and extrasolar planetary systems, especially
the magnetospheric processes in Jupiter and Saturn and the
photoevaporative processes in comets, will be addressed.
\item Stars in the Milky Way -- The studies concerning our
Galaxy will focus on various subjects: stellar clusters, for
which an UV atlas will be constructed and information about
proper motions and stellar evolution will be obtained;
planetary nebulae, their proper motions, expansion and shocks,
as well as the search for companions in infrared-bright
asymptotic-giant-branch and proto-planetary nebulae 
\citep{Gue10,Gue12}; protostellar jets and the mechanisms
at work in these interesting phenomena \citep{Gom99,Cof04};
the connection between magnetic activity and mass from solar-like
stars to brown dwarfs and planets \citep{GomMar12}; peculiar
objects, such as some of the members of the $\sigma$ Orionis
cluster, which will provide information about the complete
spectrum of stellar masses in the UV from O stars down to
T~Tauri stars and brown dwarfs \citep{Cab06}; absorption
features in the stellar radiation by the atmospheres of
transiting planets \citep{Lec04,Lec10}.
\item Extragalactic astronomy -- The ISSIS instrument will
be able to observe efficiently extragalactic objects: studies
of gravitationally lensed quasars will reveal the intrinsic
variability of the nuclear continuum, as well as the effects
on the lensing galaxy \citep{Goi10}. Lyman-$\alpha$ emitters
up to redshift $z \sim 2$ not visible from Earth can be
investigated with the WSO-UV; Atlases of massive stars in
galaxies of the Local Group will be built, together with a
detailed inspection of their winds \citep{Gar11}. High-mass
X-ray binary systems and their environments will be monitored
\citep{Bla09}.
\end{itemize}

The ISSIS scientific requirements \citep{Gom12a} driving the
design are the following:

\begin{itemize}
\item High-resolution mapping of weak and nebulous sources
such as microjets or gravitational lenses.
\item Mapping of UV emission lines in extended emission 
nebulae (H~{\sc II} regions, supernovae remnants, planetary
nebulae) and jets (from protostars or compact objects).
\item Efficient spectroscopy of weak sources such as brown 
dwarfs, micro jets or star forming galaxies at low redshift
($0.5<z<1.5$).
\item Resolution of at least R$\sim$500 to study the absorption 
of stellar radiation by transiting planets, or to derive the 
terminal velocity of winds from O stars in the Local Group.
\item Enhancement of the dynamic range with coronographs or 
masks in order to map the faint emission close to bright 
sources (e.g. disks, jets, binary components).
\item Short time resolution to track the evolution of 
instabilities in disks surrounding compact sources.
\end{itemize}

\section{ISSIS design \label{issis}}

The instrument is located below the primary mirror and above
the optical bench. This location imposes additional constraints
to the design, in terms of weight and size: a maximum of 61.5~kg
on the optical bench is allowed, and the full instrument has to be
fit within a flat cylinder of height 17~cm. ISSIS is fed by the
central part of the beam but a pick-up mirror is required to fold
the beam from the telescope adding one reflection (see Fig.~1).
\begin{figure}
\begin{center}
\includegraphics*[width=0.9\textwidth,angle=0]{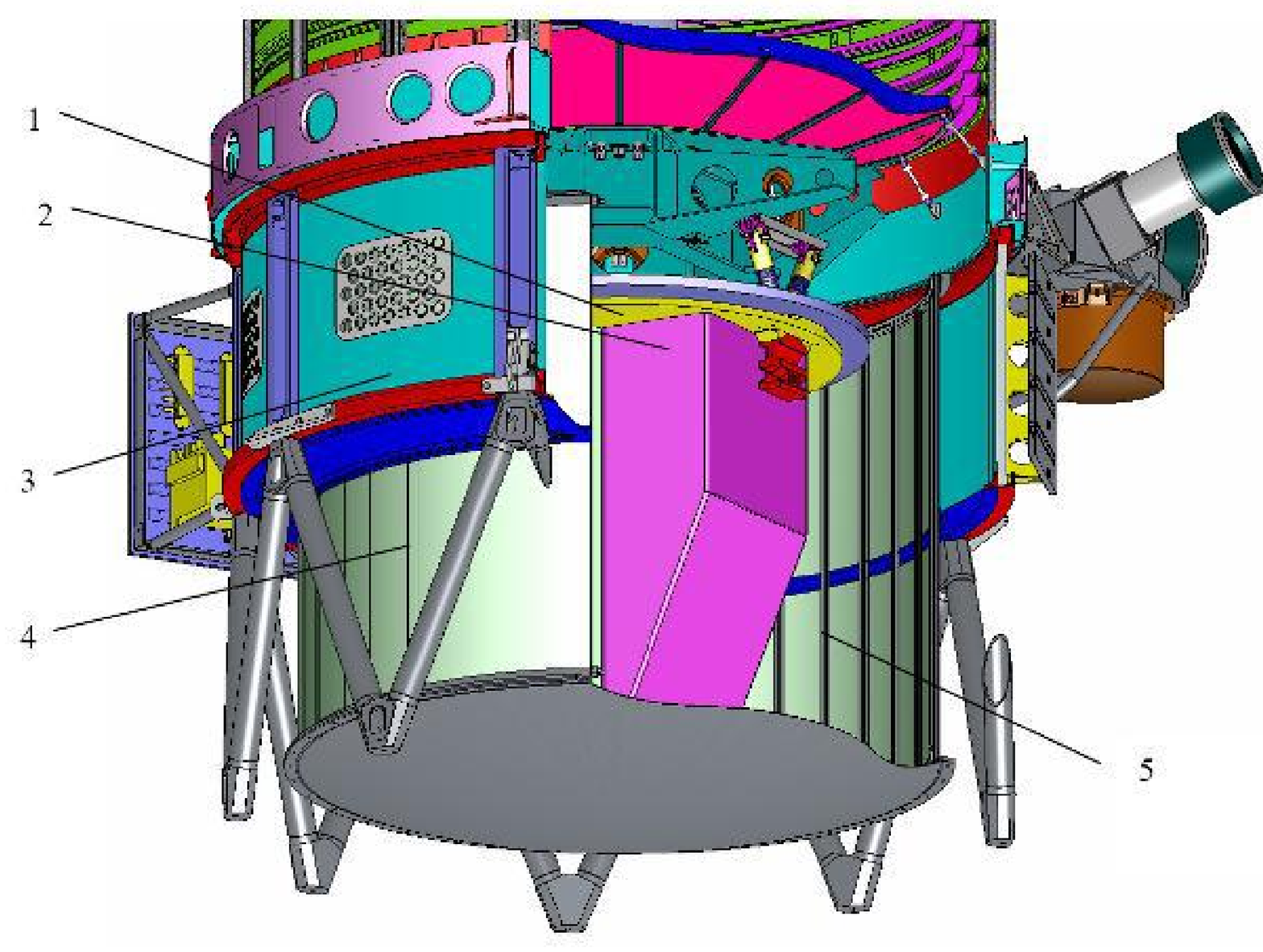}\\
\includegraphics*[width=0.9\textwidth,angle=0]{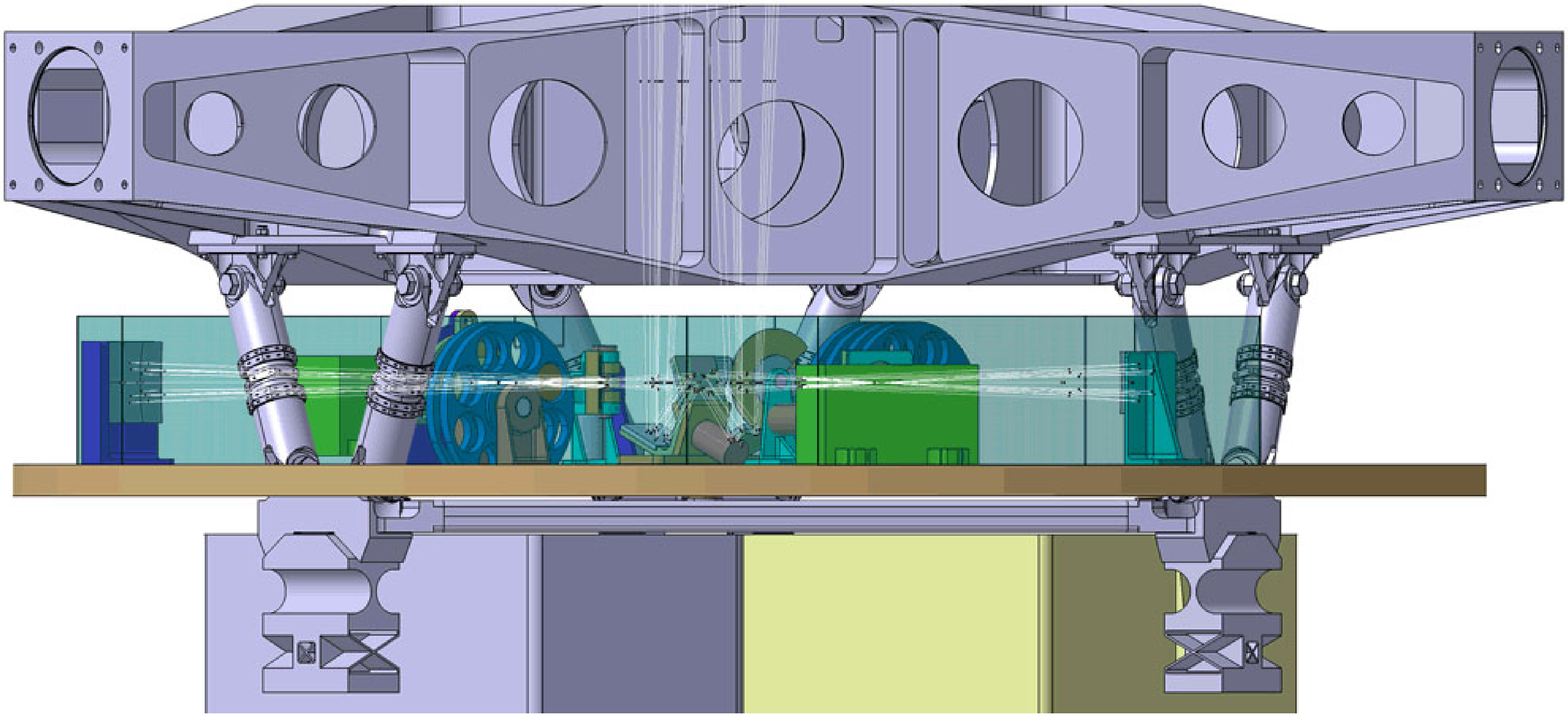}
\end{center}
\caption{{\it Top:} Instruments compartment in WSO-UV:
(1) optical bench, (2) WUVS spectrographs, (3) external
instrumentation panel, (4-5) lateral panels. ISSIS is above
the optical bench and below the primary mirror. {\it Bottom:}
Detail of the ISSIS location in the telescope.}
\end{figure}
The final design is a compromise between the scientific
requirements and the telescope/platform requirements
\citep{Gom11,Gom12a,Gom12b}.

Fig.~2 shows the layout of the instrument. 
\begin{figure}
\begin{center}
\includegraphics*[width=\textwidth,angle=0]{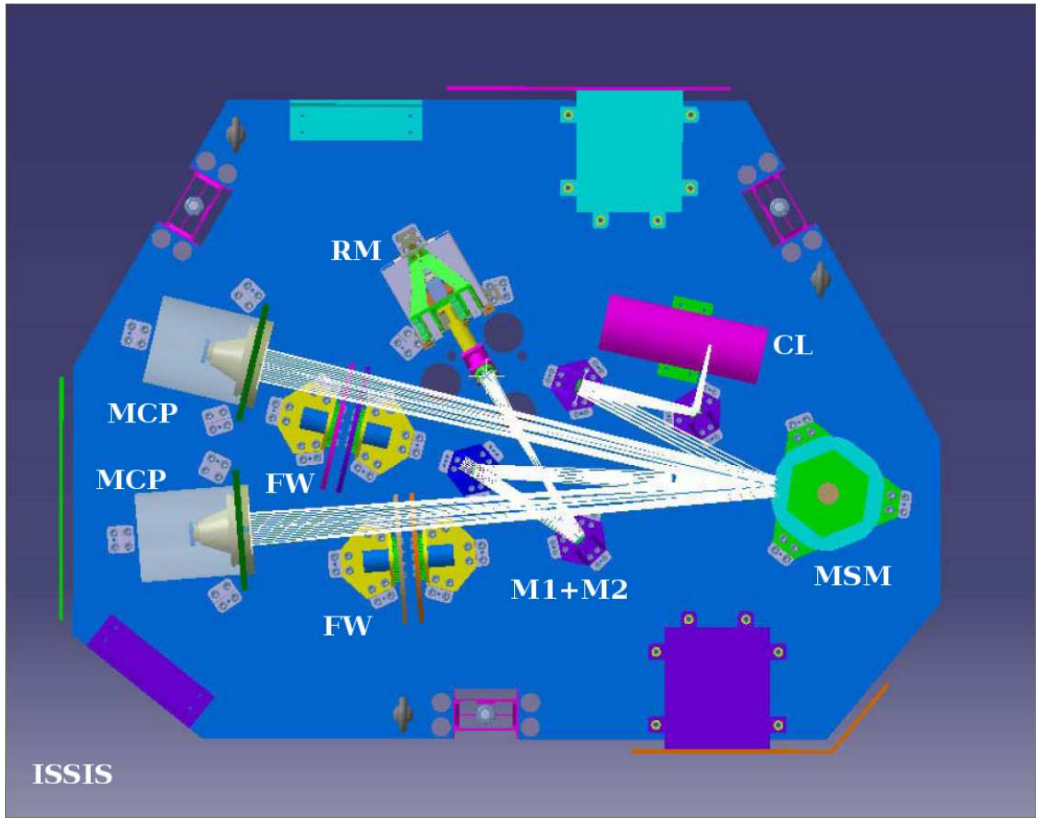}
\end{center}
\caption{Layout of the Imaging and Slitless Spectroscopy
Instrument (ISSIS). The acronyms mark the location of detectors
(MCP), filter wheels (FW), refocusing mechanism (RM) with pick
up mirror, calibration lamp (CL), mode selector mechanism (MSM)
and the mirrors M1 and M2.}
\end{figure}
ISSIS has two channels for imaging and slitless spectroscopy:
\begin{itemize}
\item The Far Ultraviolet (FUV) channel: working in the range
1150--1750~\AA.
\item The Near Ultraviolet (NUV) channel: covering wavelengths
in the 1850--3200~\AA~interval.
\end{itemize}
Both channels are equipped with photo-cathods with Micro Channel
Plate (MCP) amplifiers and CMOS reading. CsI and CsTe photocathods
have been selected for the FUV and NUV channels, respectively. The
use of MCPs imposes strong constraints to the instrument capabilities,
in particular on the field of view (FoV), barely 70~arcsec, and on the
dynamical range. There are ongoing studies to increase the dynamical
range by using masks. A set of neutral filters has been included to
be able to satisfy the sensitivity requirements keeping the possibility
to observe sources of moderate brightness. Slitless spectroscopy is not
only an efficient mean to analyse multi-objects fields, but at the same
time it acts as a narrow-band filtering technique.

The ISSIS layout is arranged as a telephoto system, where positive
elements are followed by negative elements, to enlarge the overall
focal length. A refocusing mechanism, located at the coupling stage
between the telescope and ISSIS, changes the distance between the
intermediate image from the telescope and the first mirror of 
ISSIS. The operation modes are imaging, spectroscopy, and
calibration, and are selected through a mode selection
mechanism (MSM; see Fig.~3):
\begin{figure}
\begin{center}
\includegraphics*[width=0.7\textwidth,angle=0]{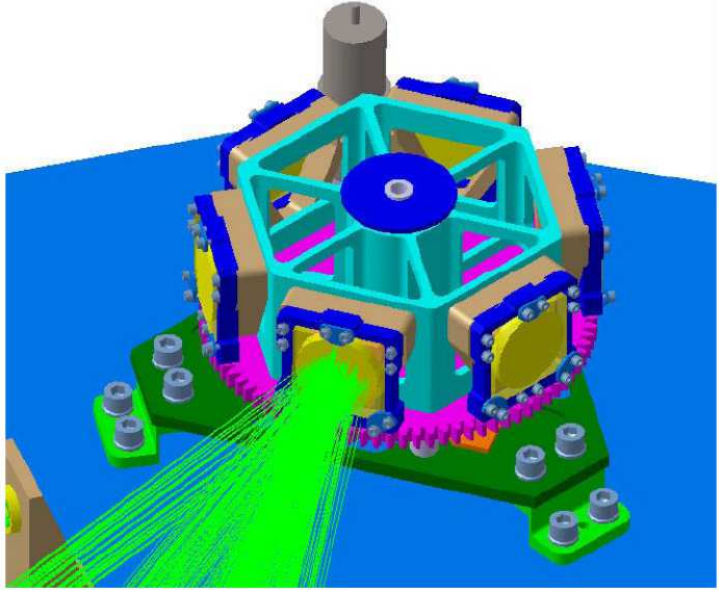}
\end{center}
\caption{Mode selection mechanism in ISSIS.}
\end{figure}
\begin{itemize}
\item Imaging: the MSM uses flat mirrors aligned to direct the
light into the selected channel (FUV or NUV). The optical filters
are accommodated on two wheels for each channel; there are
long-pass and narrow-band filters for specific investigations,
as well as two neutral filters for each channel, to increase
the dynamical range.
\item Spectroscopy: the beam is directed to the dispersive 
elements, which are reflection gratings located at the MSM.
In the nominal spectroscopy mode, the filter wheels are
positioned into a hole configuration (no filter). The FUV
diffraction grating has peak efficiency at 1400~\AA\ and 
groove density of 450~lines/mm, while the NUV one has the
peak at 2300 \AA\ and 250~lines/mm.
\item Calibration: a shutter is used to block the light
from the telescope, and the beam is directed to the calibration
subsystem in order to take flat-field images on a pixel-to-pixel
basis. The calibration unit includes the lamp, optics and shutter.
\end{itemize}

\section{ISSIS performance}

The total throughput of ISSIS includes the reflectivity of the
T-170M telescope and ISSIS mirrors, the diffraction grating
reflectivity and the sensitivity curves of the MCP detectors.
The expected value for the imaging mode will be of about 1\,\%
at 1300~\AA\ (FUV) and 7\,\% at 2500~\AA\ (NUV), while for
slitless spectroscopy the peaks will be about 0.4\,\% at
1300~\AA\ and 3\,\% at 2500~\AA\ (FUV and NUV, respectively).
The radiometric sensitivity of ISSIS can be evaluated using
the exposure time calculator (ETC) described below. Fig.~4
shows the throughput of the various components and the
estimated total throughput in the whole wavelength range.
\begin{figure}
\begin{center}
\includegraphics*[width=0.8\textwidth,angle=0]{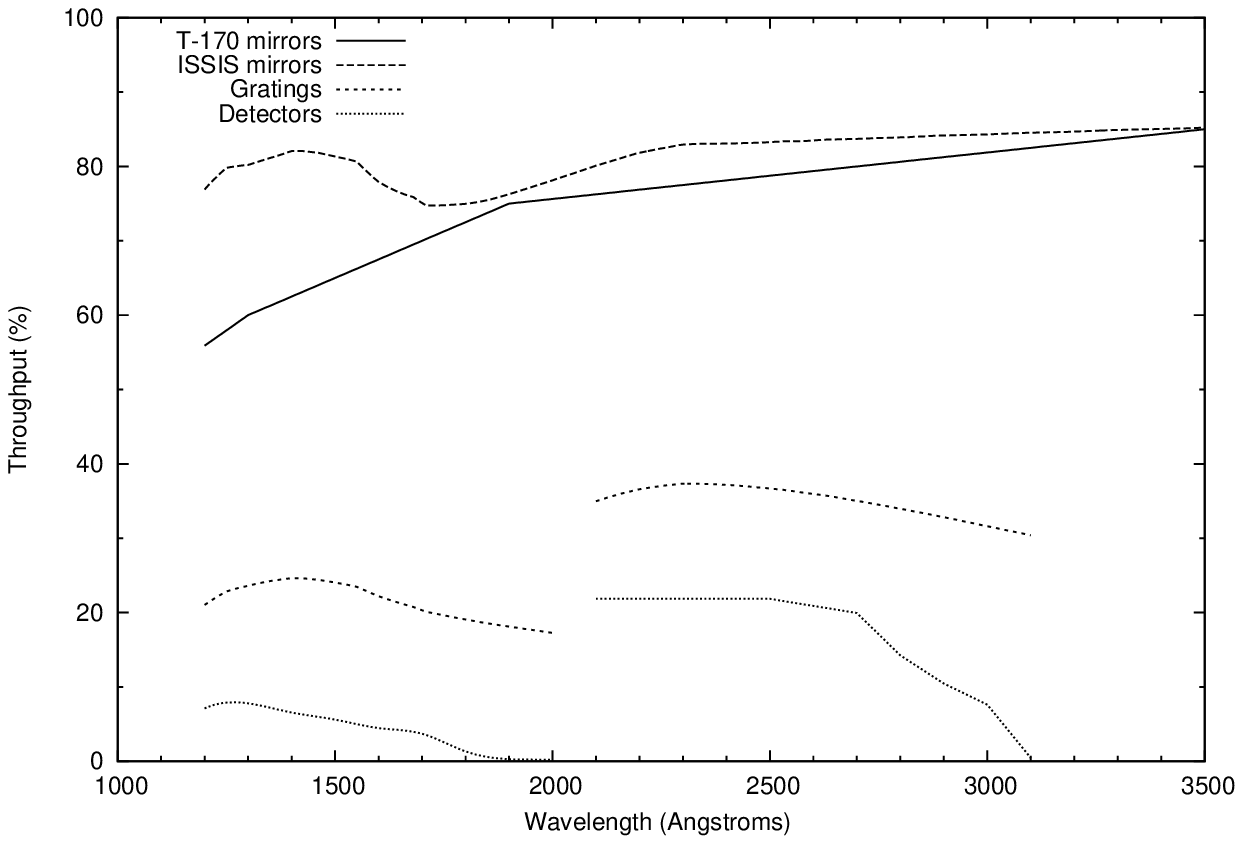} \\
\includegraphics*[width=0.8\textwidth,angle=0]{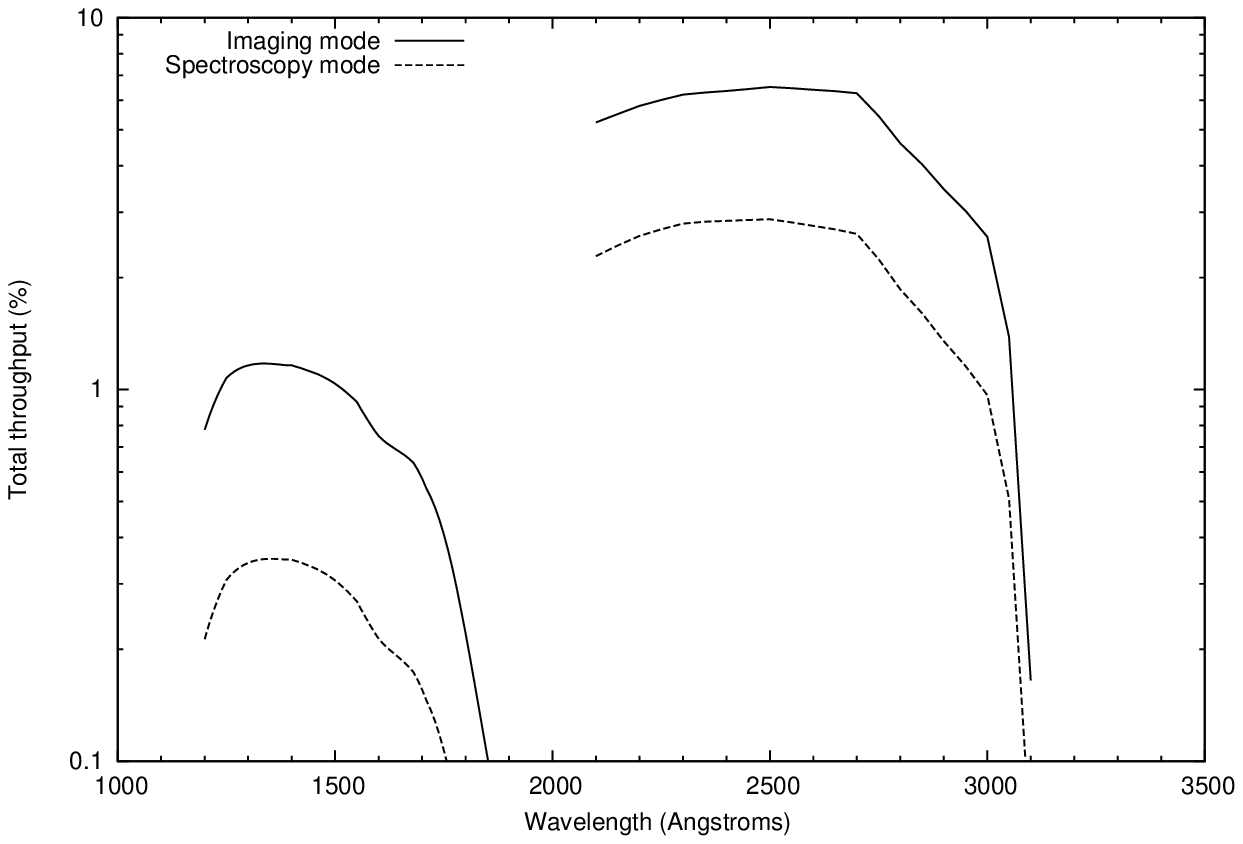}
\end{center}
\caption{{\it Top:} Sensitivity curves of the optical 
components for ISSIS: T-170 mirrors, ISSIS internal
mirrors, gratings and detectors. {\it Bottom:} total
throughput of the system in the imaging and spectroscopy
modes. Filters are not included.}
\end{figure}

One of the ISSIS key drivers is the mapping of extended
sources; this implies a proper sampling of the the point
spread function (PSF). The Fine Guiding Sensors in the
WSO-UV should provide a spatial resolution of 0.1~arcsec
at 3$\sigma$. Thus, the ISSIS optical system modifies the
T-170M focal length to have this PSF sampled by 2--3 pixels.
According to optical simulations, an asymmetrical PSF is
expected: this effect cannot be avoided at UV wavelengths,
since the optical quality is dominated by geometrical,
non-symmetric aberrations. The asymmetry in the PSF is
close to 17\,\% and to 8\,\% in the FUV and NUV channels,
respectively. Fig.~5 displays an example of PSF simulation
on two different scales; since the two channels share the
same optics, a similar PSF behaviour is expected for both
of them.
\begin{figure}
\begin{center}
\includegraphics*[width=\textwidth,angle=0]{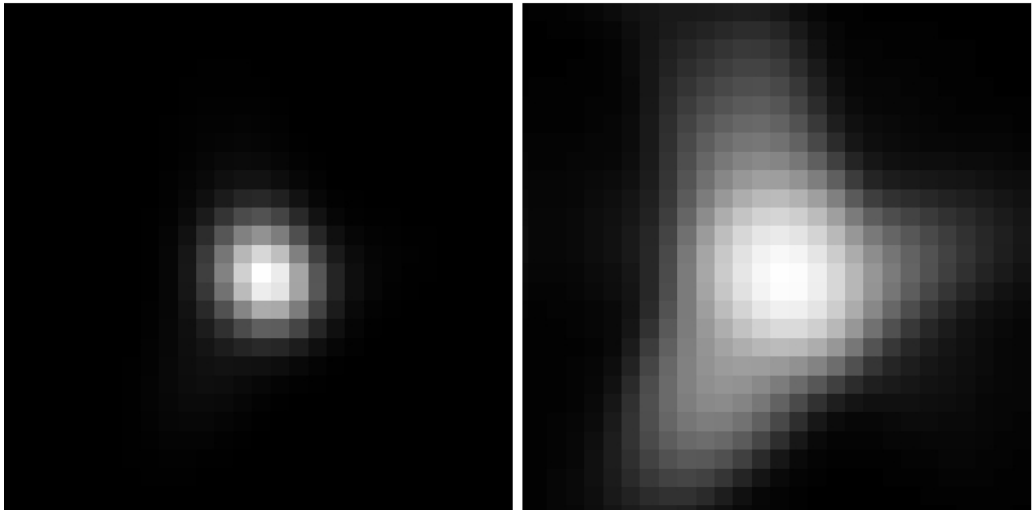}
\end{center}
\caption{A simulation of the PSF for the FUV channel in
imaging mode in linear (left) and logarithmic (right)
scales.}
\end{figure}

The field of view is limited by the photocathods sensitive
area in the MCP detectors and the spatial resolution to
$\sim$1.2~arcmin. This field of view will permit direct
imaging as well as obtaining images of extended nebular
objects in the most prominent spectral lines in the 
slitless spectroscopy mode. Table~\ref{tabISSIS} shows
a summary of the main ISSIS characteristics.
\begin{table}[ht]
\caption{Characteristics and performance of the ISSIS instrument.}
\begin{tabular}{lcc} 
\hline 
& FUV channel & NUV channel \\
\hline 
Spectral range           & 1150--1750 \AA & 1850--3200 \AA \\
FoV imaging              & $70\times75$ arcsec$^2$ & $70\times75$ arcsec$^2$\\
FoV spectroscopy         & $36\times65$ arcsec$^2$ & $31\times61$ arcsec$^2$\\
Pixel scale              & 0.036 arcsec & 0.036 arcsec \\
Scale ratio              & $<$ 7\,\% & $<$ 7\,\% \\
Number of reflections    & 4 & 4 \\
Temporal resolution      & 40 ms & 40 ms \\
Detector type            & CsI MCP & CsTe MCP \\
Detector diameter        & 40 mm & 40 mm \\
Peak throughput          & $\sim$ 1300 \AA & $\sim$ 2500 \AA \\
Spectroscopy resolution  & R $=$ 500 & R $=$ 500 \\
Spatial resolution       & 0.11 arcsec & 0.11 arcsec \\
Detector format (equivalent) & $>$ $2048\times2048$ pixels &
                               $>$ $2048\times2048$ pixels\\
\hline
\end{tabular} 
\label{tabISSIS} 
\end{table}

\subsection*{Exposure time calculator}

Astronomers can use the ETC developed by the science team at
Universidad Complutense de Madrid to evaluate the exposure
time needed for an observation, in order to achieve a desired
signal-to-noise ratio (S/N), and vice versa. The ETC is a
web-based application available through the project website
at \href{http://www.wso-uv.es}{http://www.wso-uv.es}. It
appears as a user-friendly interface with five main fields
shown in Fig.~6, which are:
\begin{figure}
\begin{center}
\includegraphics*[width=\textwidth,angle=0]{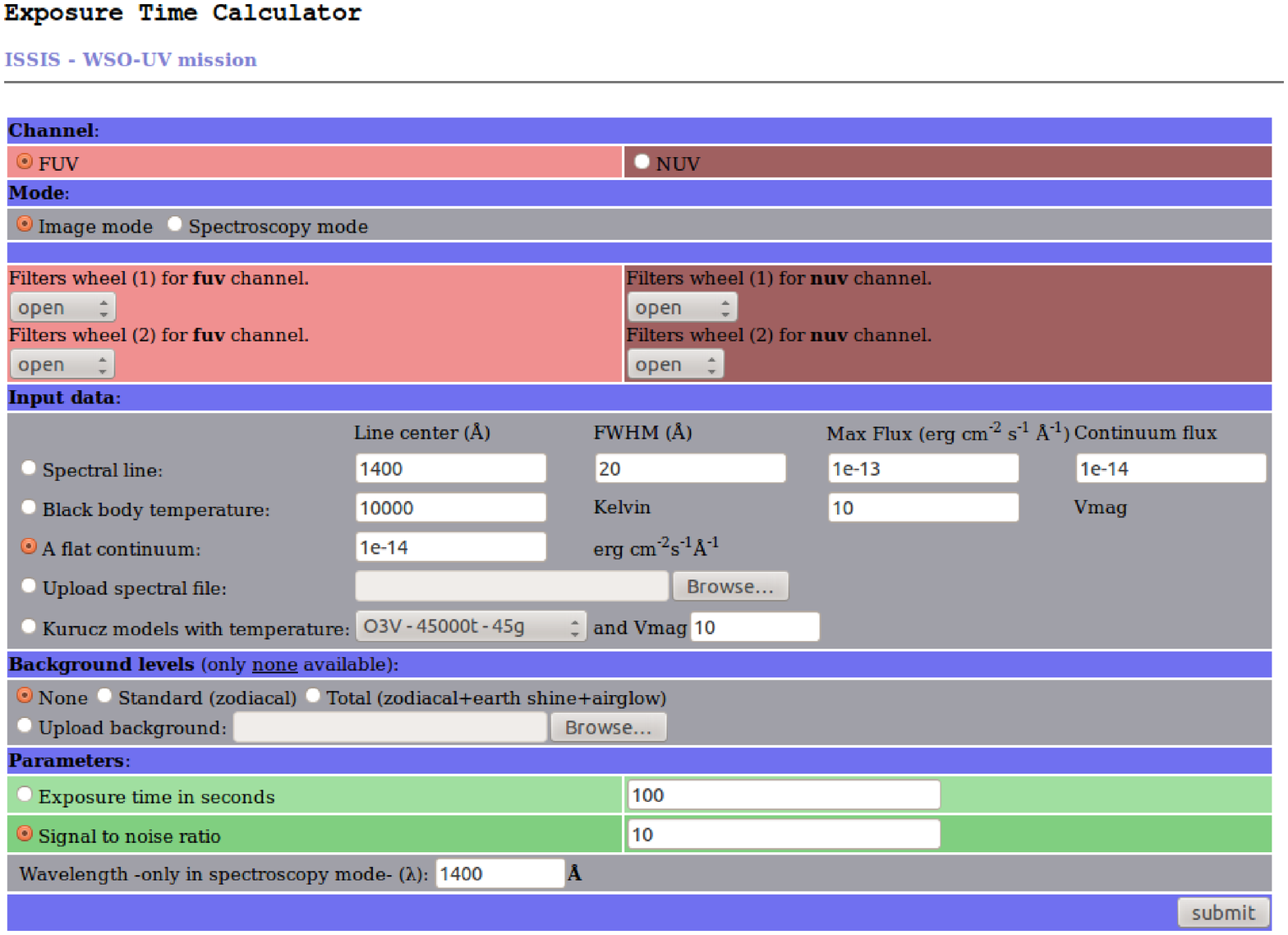}
\end{center}
\caption{Graphic interface of the ISSIS exposure time calculator.}
\end{figure}
\begin{itemize}
\item Channel: to select among the two ISSIS channels (FUV and NUV).
\item Mode: for each channel the spectroscopy or imaging mode are
allowed, and the user can select the optical filters on the wheels.
\item Input data: the input can be given as a continuum flux, 
setting its value, a spectral line (absorption or emission), 
or a predefined distribution uploaded by the user as an ASCII file.
Other options are a blackbody distribution (by choosing the effective
temperature and magnitude), or an ATLAS9 atmosphere model by \citet{Cas04}.
\item Parameters: in this section the user gives the desired S/N, and
obtains the exposure time from the calculations, or vice versa.
\end{itemize}
The output window shows the calculated S/N or time, depending
on the user's choice, and three plots showing the input spectral
distribution, the system throughput (telescope + ISSIS optical
system), and the final observed flux. The user is allowed to
see the results as a table, either in ASCII or HTML format.

\section{Conclusions}

This article presents ISSIS design at the end of Phase B. The
instrument will be a versatile tool to analyse the UV sky that
complements the other instruments onboard the WSO-UV space
telescope. ISSIS flight model is due to be deliver to the
Russian Federal Space Agency (Roscosmos) in 2015 for final
integration and assembly. The WSO-UV launch is foreseen for
2017.

\section*{Acknowledgements}

ISSIS development is being funded by Ministry of Industry,
Tourism and Commerce of Spain. We acknowledge the Spanish
Science Team for their support in this project. 
We also would like to thank to the referees for their
detailed revision of this manuscript.
The science team at Universidad Complutense de Madrid
acknowledges the financial support of the Ministry of
Economy and Competitivity through grants
AYA2008-06423-C03-01 and AYA2011-29754-C03-C01.

\end{document}